\documentclass[12pt]{article}
\usepackage{amsmath} 
\usepackage{graphicx}
\usepackage{subfigure}
\usepackage{float}
\newcommand{\goodgap}{
\hspace{\subfigtopskip}
\hspace{\subfigbottomskip}}
\newcommand{\e}[1]{\textrm{e}^{#1}}
\begin{document}
\title{THE DYNAMICAL BEHAVIOUR OF TEST PARTICLES IN A QUASI--SPHERICAL  SPACETIME AND THE PHYSICAL MEANING OF SUPERENERGY}
\author{L. Herrera$^{1}$\thanks{e-mail: laherrera@cantv.net.ve}, J.
Carot$^{2}$\thanks{e-mail: jcarot@uib.cat},
N. Bolivar$^{1}$\thanks{e-mail: nbolivar@fisica.ciens.ucv.ve}, and E. Lazo$^{1}$\thanks{e-mail: emiliolazozaia@gmail.com}\\
\small{$^1$Escuela de F\'{\i}sica, Facultad de Ciencias,} \\
\small{Universidad Central de Venezuela, Caracas, Venezuela.}\\
\small{$^2$Departament de  F\'{\i}sica,}\\
\small{Universitat Illes Balears, E-07122 Palma de Mallorca, Spain}\\
}
\maketitle

\vspace{-0.5cm}

\begin{abstract}
We calculate  the instantaneous proper radial acceleration of test particles  (as
measured  by a locally defined Lorentzian observer) in a Weyl
spacetime, close to the horizon. As expected from the Israel
theorem, there appear some bifurcations with respect to the
spherically symmetric case (Schwarzschild) which are explained in
terms of the behaviour of the superenergy, bringing out the
physical relevance of this quantity in the study of general
relativistic systems.
\end{abstract}

\section{Introduction}
As it is well known, since the seminal paper by Israel  \cite{1},
the only static and asymptotically-flat vacuum space-time possessing
a regular horizon is the Schwarzschild solution. All the other Weyl
exterior solutions \cite{2}, exhibit singularities in the physical
components of the Riemann tensor  at $r=2M$.

For not particularly intense gravitational fields and small
fluctuations off spherical symmetry,  deviations from spherical
symmetry may be described as perturbations of the spherically
symmetric exact solution \cite{Letelier}.

However, such perturbative scheme will eventually fail in regions
close to the horizon (although strictly speaking the term
``horizon'' refers to the spherically symmetric case, we shall use
it when considering the $r=2M$ surface, in the case of small
deviations from sphericity). Indeed,  as we approach the horizon,
any finite perturbation of the Schwarzschild spacetime becomes
fundamentally different from the corresponding exact solution
representing the quasi--spherical spacetime,  even if the latter is
characterized by parameters whose values are arbitrarily close to
those corresponding to Schwarzschild metric  \cite{4}. This, of
course,  is just an expression of the Israel theorem  (for observational differences between black holes and naked singularities see \cite{VE}, \cite{VK} and references therein).

Therefore, for strong gravitational fields, no matter how small the
multipole moments of the source are (those higher than monopole),
there exists a bifurcation between the perturbed Schwarzschild
metric and all the other Weyl metrics (in the case of gravitational
perturbations).

Examples of such a  bifurcation have been brought out in  the study
of the trajectories of test particles in the $\gamma$ spacetime
\cite{zipo}, and  in the M--Q spacetime \cite{yo},\cite{yobis}, for
orbits close to $2M$ \cite{HS},\cite{herreramq}.

Also, the influence of the quadrupole moment on the motion of test
particles within the context of Erez--Rosen metric \cite{erroz} has
been investigated by many authors (see \cite{quevedo1} and
references therein).

The purpose of this paper is to explain the bifurcation mentioned
above, in terms of the behaviour of super--energy \cite{Bel} in a
neighborhood of the horizon. This quantity, which may be defined
from the Bel \cite{bel1} or the  Bel--Robinson tensor \cite{bel2}
(they both coincide in vacuum), has been shown to be very useful
when it comes to explaining  a number of  phenomena in the context of
general relativity.

Thus, for instance, it helps to explain the occurrence of vorticity
in  both  radiative  \cite{HB}, and stationary spacetimes \cite{HC}.
Also, it renders intelligible the behaviour of test particles moving
in circles around the symmetry axis in an Einstein--Rosen spacetime
\cite{HE}.

In this paper we shall see how the behaviour of  the instantaneous radial
acceleration of a test particle (as measured by a locally
defined Lorentzian observer) in a specific spacetime of the Weyl
family and in regions close to the horizon, becomes intelligible
when contrasted with the corresponding behaviour of superenergy.

The Weyl metric to be considered here is  the  M--Q spacetime. The
rationale for this choice is that  due to its relativistic multipole
structure, the M--Q solution (more exactly, a sub--class of this
solution M--Q$^{(1)}$, \cite{yo}) may be interpreted as a quadrupole
correction to the Schwarzschild space--time, and therefore
represents a good candidate among known Weyl solutions, to describe
small deviations from spherical symmetry.

For this metric we shall calculate the instantaneous radial  acceleration of
a test particle and the superenergy. Then the very peculiar
behaviour of the former (close to the horizon) will be be explained
in terms of the behaviour of the latter.

The paper is structured as follows: in section 2 we present the Weyl
family of metrics we shall be concerned with and briefly discuss
some of its properties; next, in section 3, we calculate the 
radial four--acceleration of test particles in such setup.  In section 4 we review and
discuss the concept of superenergy.  In section 5 we briefly describe the M--Q solution and particularize the expressions for the proper
radial four--acceleration and the superenergy  for 
the case of the  M--Q$^{(1)}$ metric. Finally results are discussed in last section.

\section{The Weyl metrics}

Static axisymmetric solutions to Einstein's equations are given by
the Weyl metric \cite{2}
\begin{equation}
ds^2 = \e{2 \Psi} dt^2 - \e{-2 \Psi} [\e{2 \Gamma}(d \rho^2 +dz^2)+\rho^2
d \phi^2 ],
\label{elin}
\end{equation}
For vacuum spacetimes, Einstein's Field Equations imply for the
metric functions
\begin{equation} \Psi_{, \rho \rho}+\rho^{-1}
\Psi_{, \rho}+\Psi_{, zz} = 0, \label{meq1}
\end{equation}
and
\begin{equation}
\Gamma_{, \rho}= \rho (\Psi_{, \rho}^2-\Psi_{, z}^2) ; \qquad
\Gamma_{, z}= 2 \rho \Psi_{, \rho} \Psi_{, z}. \label{meq2}
\end{equation}

Notice  that  (\ref{meq1}) is just the Laplace equation for $\Psi$
(in  2--dimensional Euclidean space); furthermore, it is precisely
the integrability condition for (\ref{meq2}), that is: given $\Psi$,
a function $\Gamma$ satisfying (\ref{meq2}) always exists. Since in
the weak field limit $\Psi$ is related to the Newtonian
gravitational potential, this result may be stated as saying that
for any ``Newtonian'' potential there always exists a specific Weyl
metric, a well known result.

An interesting way of writing the general solution of (\ref{meq1},
\ref{meq2}) was obtained by Erez-Rosen \cite{erroz} and Quevedo
\cite{quev}, using prolate spheroidal coordinates, which are defined
as follows
\begin{equation}
\begin{aligned}
x & = & \frac {r_{+}+r_{-}}{2 \sigma}  , \qquad y  =  \frac
{r_{+}-r_{-}}{2 \sigma} \\
r_{\pm} & \equiv & [\rho^2+(z\pm \sigma)^2]^{1/2} \\
x & \geq & 1 \quad , \quad -1 \leq y \leq 1 ,
\label{pro}
\end{aligned}
\end{equation}
where $\sigma$ is an arbitrary constant which will be identified later with
the Schwarzschild's mass.
The prolate coordinate $x$ represents a radial coordinate, whereas the
other coordinate, $y$ represents the cosine function of the polar angle.

In these prolate spheroidal coordinates, $\Psi$ takes the form
\begin{equation}
\Psi = \sum_{n=0}^{\infty} (-1)^{n+1} q_n Q_n(x) P_n(y),
\label{propsi}
\end{equation}
where $P_n(x)$ and $Q_n(y)$ are the Legendre functions of first and
second kind respectively, and $q_n$ a set of arbitrary constants.
The corresponding expression for the function $\Gamma$, may be found
in \cite{quev}.

\section{The  radial acceleration of test particles}
In order to find an expression for the instantaneous radial acceleration of test particles, it is useful to start from the geodesic equations.

These can be derived from the Lagrangian
\begin{equation}
2{\cal L}=g_{\alpha\beta}\dot{x}^\alpha\dot{x}^\beta,
\end{equation}
where the dot denotes differentiation with respect to an affine
parameter $s$, which for timelike geodesics coincides with the
proper time. Then, from  Euler-Lagrange equations it follows,
\begin{equation}
\frac{d}{ds}\left(\frac{\partial{\cal
L}}{\partial\dot{x}^\alpha}\right)-\frac{\partial{\cal L}}
{\partial x^\alpha}=0,
\end{equation}
we shall not need  the full set of geodesic equations, therefore we shall display only the one involving radial acceleration
\begin{equation}
2\ddot r g_{rr}+2\dot r(\dot r g_{rr,r}+g_{rr,\theta} \dot
\theta)-\dot t^2 g_{tt,r}-\dot r^2 g_{rr,r}-\dot \theta^2 g_{\theta
\theta,r}-\dot \phi^2 g_{\phi \phi,r}=0, \label{ere}
\end{equation}
where, instead of cylindrical coordinates $(\rho,z)$, we found
useful to work with  Erez-Rosen coordinates $(r, \theta)$  given by:
\begin{equation}
\begin{aligned}
z &=(r-M) \cos \theta \\[0.2cm]
\rho &= (r^2-2 M r)^{1/2} \sin\theta
\end{aligned}
\label{(28)}
\end{equation}
which are related to prolate coordinates, by
\begin{equation}
\begin{aligned}
x &= \frac{r}{M}-1 \\[0.2cm]
y &= \cos\theta
\end{aligned}
\label{(29)}
\end{equation}

Since we are concerned only with timelike geodesics,  the range of
our coordinates is:
$$\infty >t\geq 0  \qquad r> 2M  \qquad \pi \geq \theta \geq 0  \qquad 2\pi \geq \phi \geq 0.$$

Let us now consider the motion of a test particle along a radial geodesic, for an arbitrary value of $\theta$. 
Thus putting $\dot \theta=\dot \phi=0$ in (\ref{ere}), and using the constraint (for radial geodesics)
\begin{equation}
1=g_{tt}\dot t^2+g_{rr}\dot r^2,
\label{nc}
\end{equation}

we obtain
\begin{equation}
2 \ddot r g_{tt}g_{rr}+\dot r^2(g_{rr} g_{tt} )_{,r}- {g_{tt,r}} =0,
\label{radial1}
\end{equation}

It should be kept in mind that we are not interested in  a full  description of  the motion of test particles  (we are not going to integrate the full set of geodesic equations), but only in the expression for the radial acceleration of a test particle at any given time. Accordingly we shall not need to take into consideration the constraints impossed on $\dot r$, $\ddot \theta$ and so on,  from the the other geodesic equations, for the radial motion.

In order to express our results in terms of physically meaningful
quantities, let us introduce (locally defined) coordinates
associated with a locally Minkowskian observer (or alternatively, a
tetrad field associated with such a Minkowskian observer). Thus, let
\begin{equation}
dX=\sqrt{-g_{rr}}dr
\label{x}
\end{equation}
and
\begin{equation}
dT=\sqrt{g_{tt}}dt,
\label{t}
\end{equation}
It then follows that
\begin{equation}
\dot r=\frac{\frac{dX}{dT}}{\sqrt{g_{rr}\left[\left(\frac{dX}{dT}\right)^2-1\right]}}
\label{radial6}
\end{equation}
and
\begin{equation}
\frac{d^2 X}{dT^2}=\ddot
r\sqrt{-g_{rr}}\left[1-\left(\frac{dX}{dT}\right)^2\right]^2-
\left(\frac{dX}{dT}\right)^2
\frac{g_{rr,r}\left[1-\left(\frac{dX}{dT}\right)^2\right]}{2(-g_{rr})^{3/2}}.
\label{radial7}
\end{equation}
In the spherically symmetric case, (\ref{radial7}) reduces to
\begin{equation}
\frac{d^2 X}{dT^2}=-\frac{M}{r^2}\left[1-\left(\frac{dX}{dT}\right)^2\right] \left(1-\frac{2M}{r}\right)^{-1/2},
\label{ss}
\end{equation}
or, introducing the variable  $R=\displaystyle{\frac{r}{M}}$
\begin{equation}
\frac{d^2 X}{dT^2}=-\frac{1}{MR^{3/2}}\left[1-\left(\frac{dX}{dT}\right)^2\right](R-2)^{-1/2},
\label{ssR}
\end{equation}
which is a known result. Since ${dX}/{dT}$ is always smaller than
one, the attractive nature of gravity for any value of $r$ (larger
than $2M$) is clearly exhibited in (\ref{ss}).

\section{Superenergy}
As it is known, in classical field theory, energy is a quantity
defined in terms of potentials and their first derivatives. In
General Relativity however, it is impossible to construct a tensor
expressed only through the metric and their first derivatives (the
equivalence principle). Accordingly, a local description of
gravitational energy in terms of true invariants (tensors of any
rank) is  not possible within the context of the theory.

Thus, one is left with the following three alternatives:
\begin{itemize}
\item  Looking for a non--local  definition of energy
\item  Finding a definition based on  pseudo--tensors
\item  Resorting to a succedaneous definition, e.g.:  superenergy.
\end{itemize}

In this work we are going to explore the last alternative.  As
indicated in the Introduction, the motivations for  doing so are
given by  the rich and  profound physical meaning of such quantity.

Superenergy $W$  may be defined  from either  the Bel or the
Bel--Robinson tensor \cite{ parrado}. Since we are working with
vacuum spacetimes both definitions coincide, and one then has:

\begin{equation}
W=E^{\alpha \beta}E_{\alpha \beta}+B^{\alpha \beta}B_{\alpha \beta}
\label{s1}
\end{equation}
with
\begin{equation}
E_{\alpha \beta}=C_{\alpha \gamma \beta \delta}u^{\gamma}u^{\delta}
\label{electric}
\end{equation}
\begin{equation}
B_{\alpha \beta}=^{*}C_{\alpha \gamma \beta \delta}u^{\gamma}u^{\delta}=
\frac{1}{2}\eta_{\alpha \gamma \epsilon \rho} C^{\epsilon
\rho}_{\quad \beta \delta} u^{\gamma}
u^{\delta},
\label{magnetic}
\end{equation}
where $C_{\alpha \gamma \beta \delta}$ is the Weyl tensor,
$\eta_{\alpha \beta \gamma \delta}$ is the Levi--Civita tensor and
$u^\alpha$ is the four--velocity of observers at  rest in the frame
of  (\ref{elin}), i.e.
\begin{equation}
u^\alpha=\left(\frac{1}{\sqrt{g_{00}}},0,0,0\right).
\label{vel.}
\end{equation}

Observe that since we are working with static
spacetimes, the magnetic part of the Weyl tensor ($B_{\alpha
\beta}$)  vanishes identically.

Let us next briefly introduce the  metric we shall consider here,
and calculate  the corresponding expressions for the radial  
acceleration of a test particle and the superenergy. \vskip 1cm

\section {The Monopole--Quadrupole solution, $M-Q$}

\vskip 2mm

In \cite{yo, yobis} it was shown that it is possible to find a metric  of the
Weyl family, such that the resulting solution possesses only monopole and
quadrupole moments (in the Geroch sense \cite{ger}). The obtained solution
(M--Q)
may be written as follows:
\begin{equation}
\Psi_{M-Q}=\Psi_{q^0}+q \Psi_{q^1}+q^2 \Psi_{q^2}+\ldots =
\sum_{\alpha=0}^\infty q^\alpha\Psi_{q^\alpha} \quad ,
\label{(15)}
\end{equation}
where the zeroth order corresponds to the
Schwarzschild solution.
\begin{equation}
\Psi_{q^0}=-\sum_{n=0}^{\infty}\frac
{\lambda^{2n+1}}{2n+1}P_{2n}(\cos\theta) \quad ,
\label{(16)}
\end{equation}
with $\lambda \equiv M/r$ and
it appears that each power in $q$ adds a quadrupole correction to the
spherically symmetric solution. Now, it should be observed that due to the
linearity of Laplace equation,
these corrections give rise to a series of exact solutions. In other
words, the power series of $q$ may be cut at any order, and the partial
summation, up to that order, gives an exact
solution representing a quadrupolar correction to the  Schwarzschild solution.

Since we are interested in slight deviations from spherical
symmetry, we shall  consider the M--Q solution, only up to the first
order in $q$ (M--Q$^{(1)}$); with $q>0 \ (q<0)$ corresponding to an
oblate (prolate) source.

Thus, the explicit solution up to the first order, describing a
quadrupolar correction to the monopole (Schwarzschild  solution),
may be interpreted as the gravitational field outside a
quasi--spherical source, and it is given by (note a missprint in
equation (13) in \cite{yobis})
\begin{equation}
\begin{aligned}
\Psi_{M-Q}^{(1)} \equiv \Psi_{q^0}+q \Psi_{q^1} &= \frac12
\ln\left(\frac{x-1}{x+1}\right) + \frac58 q(3y^2-1)\times
\qquad \qquad \qquad \\[0.2cm]
&\times \bigg[\left(\frac{3x^2-1}4 -\frac1{3y^2-1}\right)
\ln\left(\frac{x-1}{x+1}\right) \\[0.2cm]
&\qquad \qquad \qquad \qquad -\frac{2x}{(x^2-y^2)(3y^2-1)} + \frac{3x}2 \bigg] \; ,
\label{(25)}
\end{aligned}
\end{equation}

\begin{align}
\Gamma^{(1)}_{M-Q} & \equiv \Gamma_{q^0}+q \Gamma_{q^1}+q^2 \Gamma_{q^2} =\frac12\left(1+\frac{225}{24}
q^2\right)\ln\left(\frac{x^2-1}{x^2-y^2}\right) \nonumber\\
&- \frac{15}8 q x(1-y^2)\left[1- \frac{15}{32} \left(x^2+7y^2-9x^2y^2+1 \phantom{\int} \right.\right. \nonumber \\
&- \left.\left. \frac83 \frac{x^2+1}{x^2-y^2}\right)\right]\ln\left(\frac{x-1}{x+1}\right) \nonumber\\
&+ \frac{225}{1024}q^2(x^2-1)(1-y^2)(x^2+y^2-9x^2y^2-1)
\ln^2\left(\frac{x-1}{x+1}\right) \label{(26)} \\
&- \frac{15}4 q(1-y^2)\left[1-\frac{15}{64}q(x^2+4y^2-9x^2y^2+4)\right]
\nonumber\\
&- \frac{75}{16}q^2 x^2\frac{1-y^2}{x^2-y^2} - \frac{5}{4}q(x^2+y^2)
\frac{1-y^2}{(x^2-y^2)^2} \nonumber\\
&- \frac{75}{192}q^2(2x^6-x^4+3x^4y^2-6x^2y^2+4x^2y^4-y^4-y^6)
\frac{1-y^2}{(x^2-y^2)^4}  \quad . \nonumber
\end{align}

In \cite{herreramq} it was shown that the behaviour of test particles in the M--Q$^{(1)}$ spacetime, becomes particularly strange on the symmetry axis;
i.e.: $\theta =0, \pi$, or else $y=\pm 1$ (close to the horizon). Therefore it is for that region that we are going to calculate  the proper radial acceleration of a particle on the axis
for the M--Q$^{(1)}$ spacetime (for all other regions, including the equatorial plane, the abnormal behaviour commented below is not observed \cite{herreramq}). Using (\ref{radial7}) we obtain
\begin{align}
\begin{aligned}
 \frac{d^2 X(T)}{dT^2} &= \frac{1}{8} \e{(5/4) q A(R)}
 \left( \frac{d X(T)}{dT} + 1 \right)
 \left( {\frac {d X(T)}{dT}} -1 \right)
 \sqrt{\frac{(R-2)^{-3}}{R^5}} \times \\[0.2cm]
 &\times \left[15q \ln \left( \frac {R-2}{R} \right)
 \left( {R}^{5}-5 {R}^{4}+8 {R}^{3} -4 R^2\right)
 \right. + \\[0.2cm]
 &+ q \left(30 R^4 - 120 R^3 + 130 R^2 - 20R + 20 \right) + 8 R^2 - 16R \bigg] \frac{1}{M}
\end{aligned}
\label{radial8}
\end{align}
where
\begin{equation}
 \begin{aligned}
  A(R) = \frac{1}{R (R-2)} \bigg[ 6R^3 - 18 R^2 + 8R + 4 + \\[0.2cm]
  + \ln \left( {\frac {R-2}{R}} \right) \left(3{R}^{4}-12 {R}^{3}+12 {R}^{2}\right) \bigg] \\
 \end{aligned}
\end{equation}
Since we are interested in the value of the radial acceleration for Lorentzian observers instantaneous at rest, we shall plot (\ref{radial8}) with $q=\pm 0.01$ and $\frac {d X(T)}{dT}=0$.

For this metric, the expression of the superenergy (again, on the
symmetry axis $y^2 =1$) reads
\begin{align}
W_{MQ} &= \frac{1}{1536} \e{(5/4) q A(R)} \times
\bigg[
 \left( 768 R^2 - 1152 R^3 + 576 R^4 - 96 R^5 \right) \nonumber \\[0.2cm]
 &+ q \left( -1920 R + 3360 R^2 - 6720 R^3 + 5880 R^4 \right. \nonumber \\[0.2cm]
 &\quad \left. - 1200 R^5 - 540 R^6 + 180 R^7 \right) \nonumber \\[0.2cm]
 &+ q^2 \left( 400 - 800 R + 5600 R^2 - 10000 R^3 + 22900 R^4 \right. \nonumber \\[0.2cm]
 &\quad - \left. 32400 R^5 + 22200 R^6 - 7200 R^7 + 900 R^8 \right) \nonumber \\[0.2cm]
 &+ q \ln\left(\frac{R-2}{R}\right) \left( 2880 R^3 - 5760 R^4 + 3600 R^5 \right. \nonumber \\[0.2cm]
 &\quad - \left. 360 R^6 - 360 R^7 + 90 R^8 \right) \nonumber \\[0.2cm]
 &+ q^2 \ln\left(\frac{R-2}{R}\right)
 \left( - 2400 R^2 + 7200 R^3 - 23400 R^4 + 49200 R^5 \right. \nonumber \\[0.2cm]
 &\quad - \left. 52500 R^6 + 29100 R^7 - 8100 R^8 + 900 R^9 \right) \nonumber \\[0.2cm]
 &+ q^2 \ln\left(\frac{R-2}{R}\right)^2
 \left( 3600 R^4 - 14400 R^5 + 23400 R^6 \right. \nonumber \\[0.2cm]
 &\quad - \left. 19800 R^7 + 9225 R^8 - 2250 R^9 + 225 R^{10} \right)
\bigg]
\frac{1}{M^{4} (R-2)^{6} R^{10}}
\end{align}

\newpage
\begin{figure}[H]
\centering
\subfigure[Superenergy $q>0$]{\label{W_qpositive}
\includegraphics[scale=0.3]{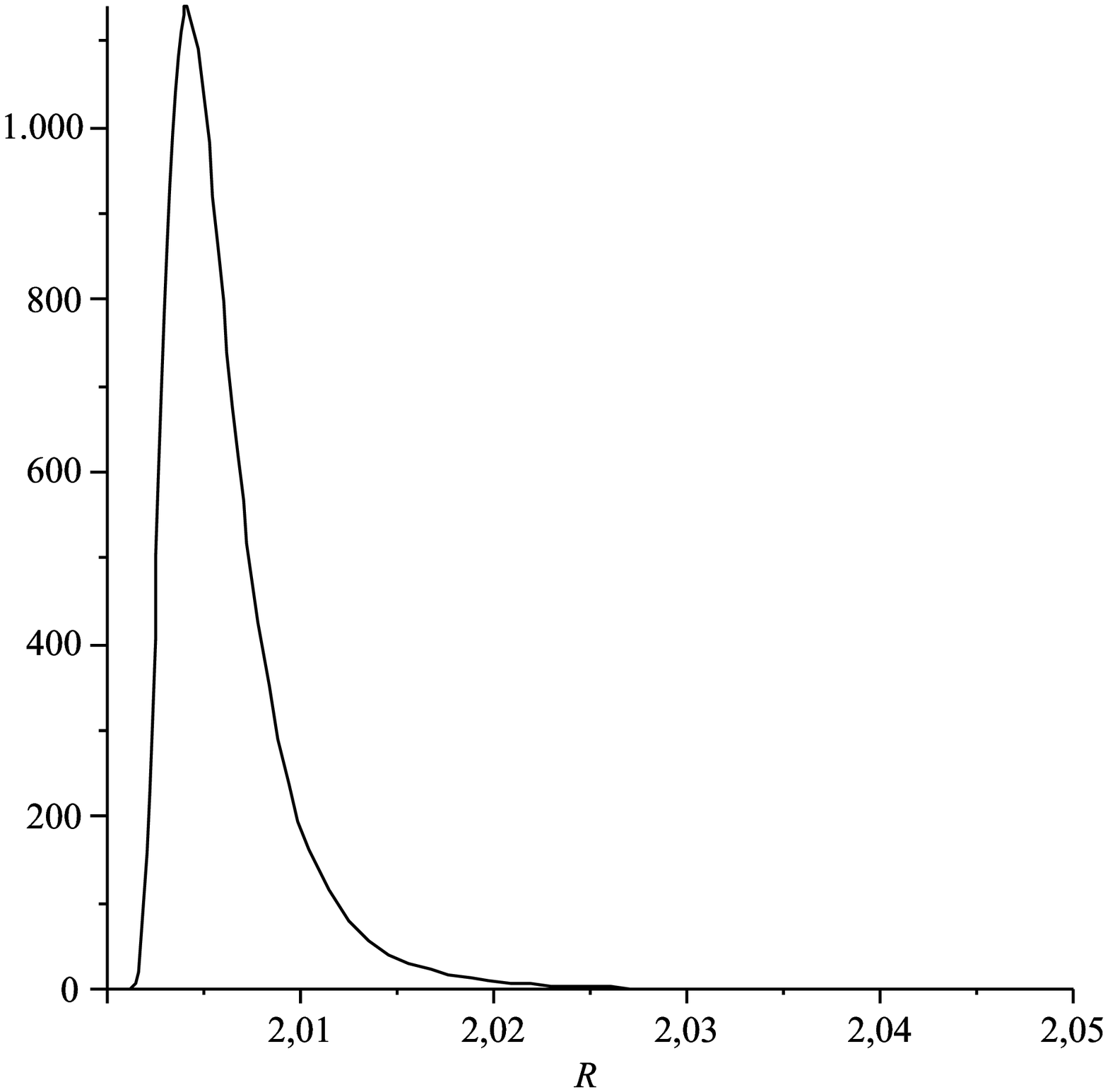}
}
\goodgap
\subfigure[Radial Acceleration $q>0$]{\label{A_qpositive}
\includegraphics[scale=0.3]{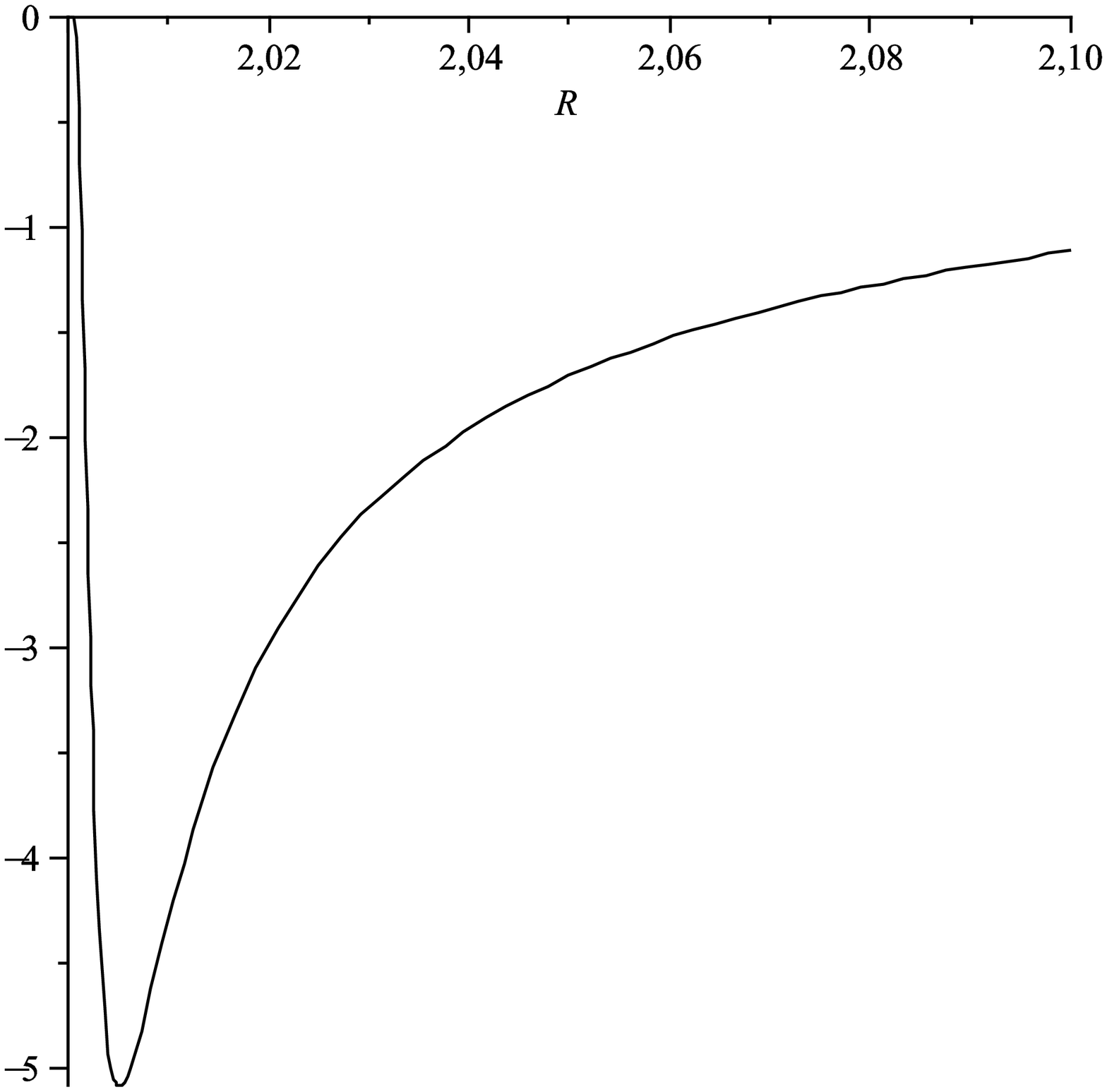}
}
\vspace{2cm}
\subfigure[Superenergy $q<0$]{\label{W_qnegative}
\includegraphics[scale=0.3]{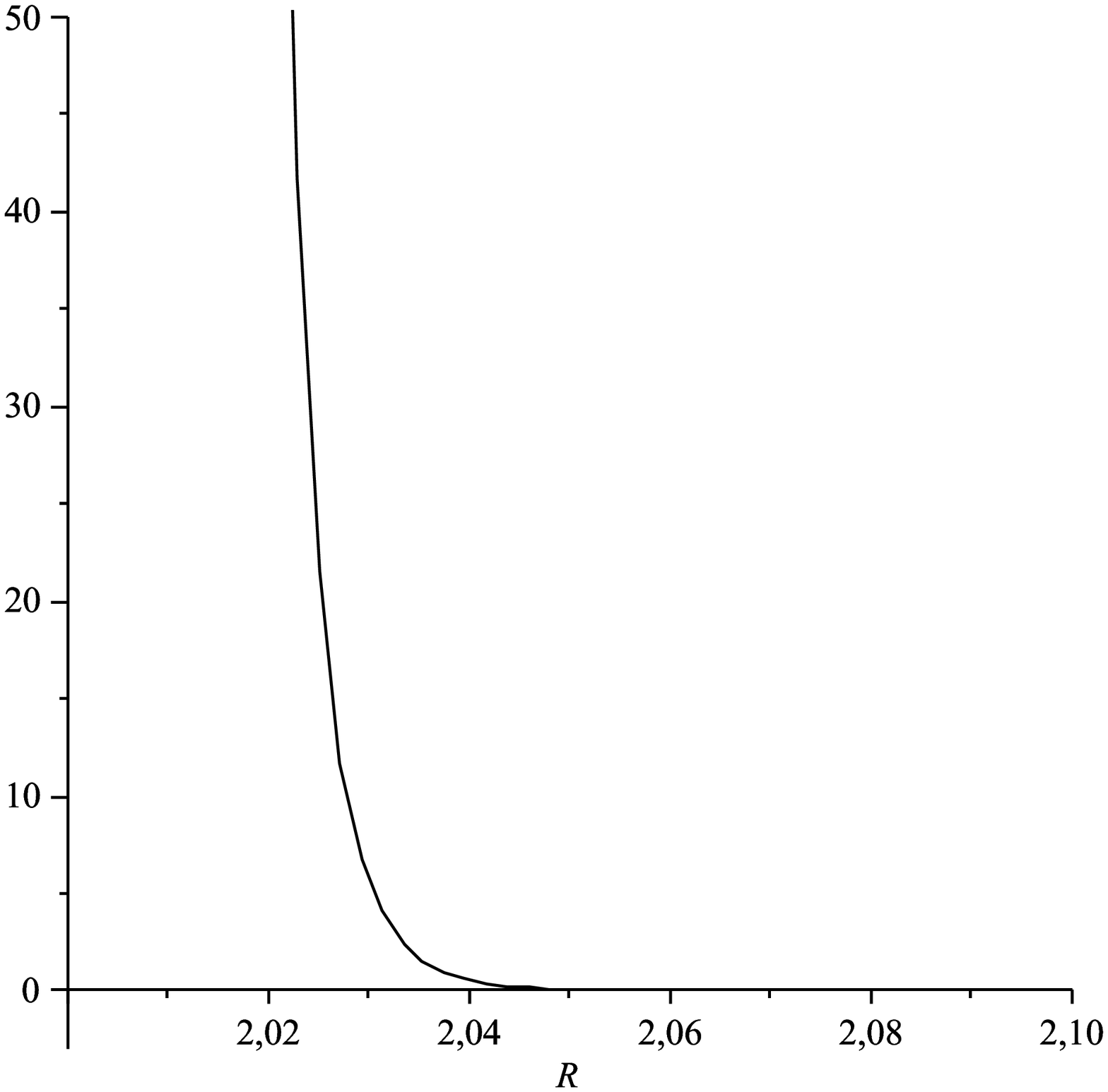}
}
\goodgap
\subfigure[Radial Acceleration $q<0$]{\label{A_qnegative}
\includegraphics[scale=0.3]{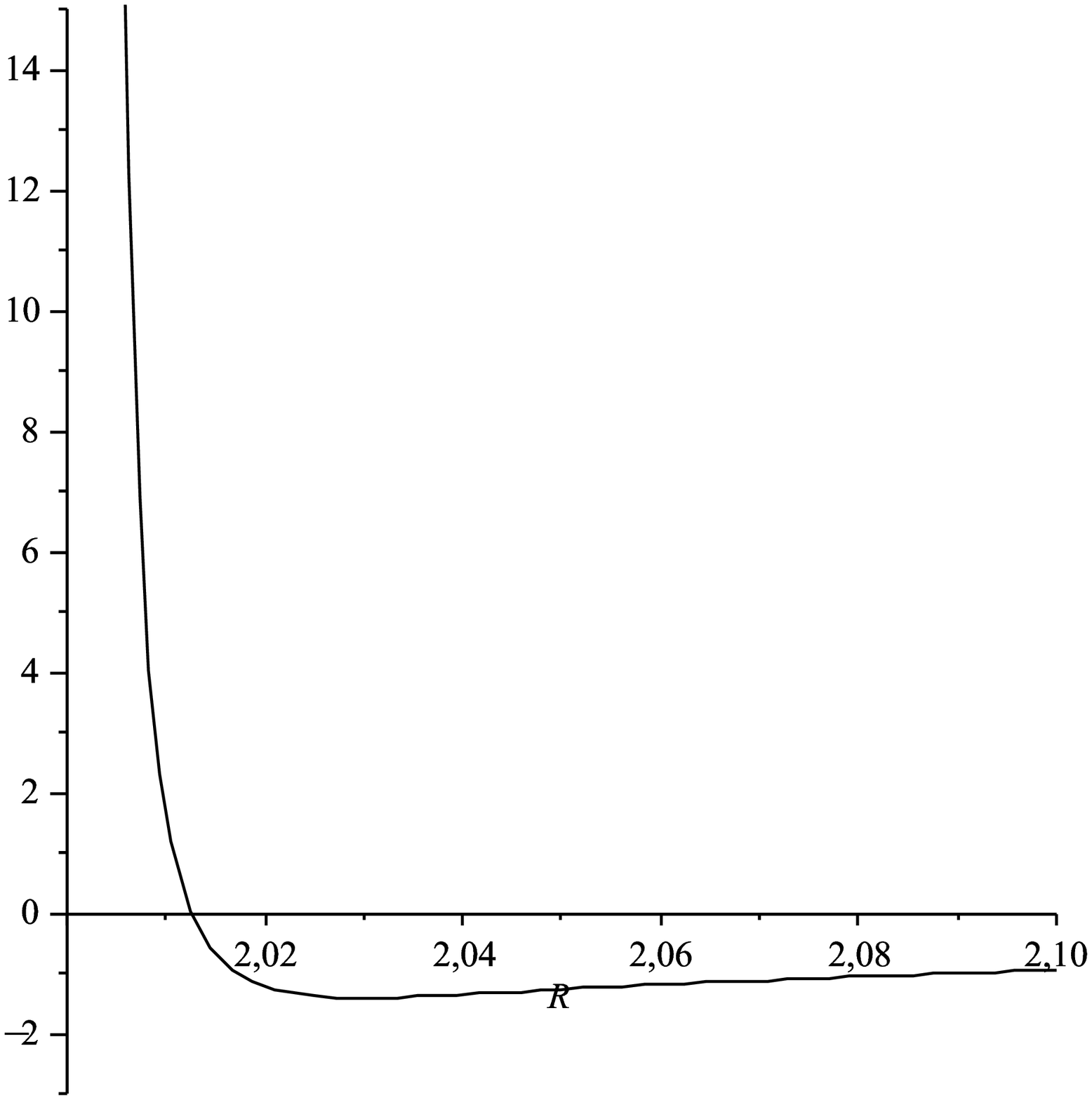}
}
\end{figure}
Figures \subref{W_qpositive} and \subref{A_qpositive} show the
behaviour of $W$ and $\frac{d^2X(T)}{dT^2}$ for $q=0.01$, whereas
figures \subref{W_qnegative} and \subref{A_qnegative} display their
behaviour for $q=-0.01$. We shall next discuss the results obtained so
far.

\newpage
\section{Discussion}
The first  conclusion which emerges from figures
\subref{W_qpositive} and \subref{A_qpositive} is that, close to the
horizon, the behaviour of the test particle (in what concerns
${d^2X(T)}/{dT^2}$) is extremely abnormal, as expected from Israel's
theorem.

Thus for the  M--Q$^{(1)}$ metric with $q>0$ it appears that a test
particle placed on the axis of symmetry in the neighbourhood of the
horizon does not  feel any attraction from the source
(${d^2X(T)}/{dT^2}\approx 0$). Still more shocking: as we move
outwards (always on the symmetry axis), the magnitude of
$\frac{d^2X(T)}{dT^2}$ increases with $R$, until some value of $R$,
from which it starts to ``behave'' properly (figure
\subref{A_qpositive}).

This pathological behaviour of ${d^2X(T)}/{dT^2}$ is  fully
consistent with that of superenergy in the same range of $R$, as
indicated in figure \subref{W_qpositive}. Indeed, $W$ also vanishes
close to the horizon, increasing as we move outwards along the
symmetry axis, until we are far away enough from the horizon and the
expected behaviour is recovered.

For $q<0$ the situation is still more unusual. Indeed, on a
neighbourhood of the horizon, on the axis of symmetry,
${d^2X(T)}/{dT^2}>0$, implying that the particle experiences a
repulsive force. This effect  is  restricted to values of $R$ very
close to $2$. As we move away from the horizon  the proper
acceleration becomes negative, although still displaying an abnormal
behaviour since it increases in magnitude with $R$. Moving further
away from the origin (along the symmetry axis)  we recover the
``normal'' behaviour (${d^2X(T)}/{dT^2}$ (negative and decreasing
with $R$). The dependence of $W$ with $R$ in this case, displayed in
figure \subref{W_qnegative}, is consistent with the graphics of
${d^2X(T)}/{dT^2}$ above. Indeed, in a neighborhood of the horizon,
$W$ is singular and so is its  derivative with respect to $R$, this
explaining the pathological behaviour of ${d^2X(T)}/{dT^2}$ in that
range of values of $R$.  As we move sufficiently far away from $R=2$
we recover the expected behaviour.

Thus we have seen that the concept of superenergy is a suitable
measure of the strength of gravitational interaction, even in highly
pathological situations. The fact that it is a true scalar (obtained
from a true tensor) reinforces further its relevance in the study of
self--gravitating systems.

\section*{Acknowledgments.}
One of us (JC) gratefully acknowledges financial support from the
Spanish Ministerio de Educaci\'{o}n y Ciencia through the grant
FPA2004-03666. (LH) wishes to thank  financial support from the
FUNDACION EMPRESAS POLAR,  Universitat de les  Illes Balears and
CDCH at Universidad Central de Venezuela under grant PG
03-00-6497-2007.

\end{document}